\documentstyle[12pt,aps]{revtex}%

\newcommand{\be}{\begin{equation}}
\newcommand{\ee}{\end{equation}}
\newcommand{\bea}{\vspace{0.25cm}\begin{eqnarray}}
\newcommand{\eea}{\end{eqnarray}}

\def\PLA{{Phys. Lett.}  A }
\def\PRL{{Phys. Rev. Lett.} }
\def\PRA{{Phys. Rev.} A }

\begin{document}

\centerline{ \large \bf A biphotons double slit experiment} 
\vskip 1.2cm
\centerline{G. Brida, E. Cagliero, G. Falzetta, M. Genovese \footnote{genovese@ien.it, http://www.ien.it/~genovese/marco.html}, M. Gramegna, }
\vskip 0.5cm
\centerline{ \it Istituto Elettrotecnico Nazionale  Galileo Ferraris,}
\centerline{ \it  Strada delle Cacce 91, 10135 Torino, Italy }
\centerline{E. Predazzi}
\centerline{\it Dip. Fisica Teorica Univ. Torino and INFN, via P. Giuria 1, 10125 Torino, Italy}
\vskip 1cm

\centerline{ \bf Abstract}
\vskip 0.2cm
 
In this paper we present a double slit experiment where two undistinguishable photons produced by type I PDC are sent each to a well defined slit.
Data about the diffraction and interference patterns for coincidences are presented and discussed. An analysis of these data allows a first test of standard quantum mechanics against de Broglie-Bohm theory.
 

\vskip 1cm

\centerline{ \bf Introduction}
\vskip 0.5cm

Double slit experiments are textbook proof  of the complementarity principle in Quantum Mechanics and have represented a very important test bench of this theory \cite{1}.

Particularly interesting examples of double slit experiments have been realised by using biphothon fields produced in parametric down conversion, a phenomenon where an incident high frequency photon is converted, inside a non-linear crystal, into a pair of highly correlated photons (usually dubbed { \it idler} and { \it signal}) fulfilling the condition, known as phase matching condition,  that the produced pair is such that the frequencies of the produced photons sum up to the frequency of the pump photon (energy conservation) and the wave vector of the pump photon is the vectorial sum of the wave vectors  of the produced photons (impulse conservation).
Among these experiments, a first class was devoted to the study of the effect on the correlations of a double slit inserted {\it in one of the paths}, i.e. on the idler or the signal photon direction \cite{class1}. A second class was addressed to study the effect of a double slit inserted on the paths of both idler and signal \cite{class2} showing the highly non-classical aspects of PDC emission. 

In our experiment we have realised  a rather different configuration, where two  degenerate identical photons produced in PDC reach a well defined slit of a double slit at the same time. As idler and signal photons have no precise phase relation \cite{int} and each photon crosses a well defined slit, no interference appears at single photon detection level. When the coincidence pattern is considered, path undistinguishability is established since the photodectector 1 (2) can be reached either by the photon which crossed slit A or by the one that went through slit B and vice versa. Thus, even if no second order interference is expected, a fourth order interference  modulates the observed diffraction coincidence pattern. 

The main result of our experiment is that  our scheme realises the configuration recently suggested by two theoretical groups \cite{4,5,6} to test the de Broglie-Bohm (dBB) theory against standard quantum mechanics (SQM). dBB \cite{7} is a deterministic theory where the hidden variable (determining the evolution of a specific system) is the position of the particle, which follows a perfectly defined trajectory in its motion. The evolution of the system is given by classical equations of motion, but an additional potential  must be included. This "quantum" potential is related to the wave function of the system and thus  is non-local. The inclusion of this term, together with an initial distribution of particle positions given by the quantum probability density,  successfully allows the reproduction of {\it almost} all the predictions of quantum mechanics. Nevertheless, a possible discrepancy between SQM and dBB in specific cases has been recently suggested by  Ref. \cite{4,5,6}.
Although this conclusion is still somehow subject to discussion \cite{disc}, we think that our results, in agreement with SQM predictions but at variance with dBB ones (see \cite{nosD} for a further discussion), represent a relevant contribution to the debate about the foundations of quantum mechanics urging a final clarification its validity. 
\vskip 1cm

\centerline{ \bf Description of the experiment}
\vskip 0.5cm

In our set-up (see Fig.1)  a 351 nm pump laser of 0.5 W power is directed into 
a lithium iodate crystal, where correlated pairs of photons are generated by 
type I Parametric Down Conversion (i.e. the two photons have the same polarisation). The pairs of photons are emitted 
at the same time within femtoseconds, whilst the correlation time is some orders of magnitude larger, and on a well defined direction for a 
specific frequency. By means of an optical condenser and within two correlated directions corresponding both to 702 nm emission (the degenerate emission for a 351 nm pump laser), the produced photons are sent 
on a double slit (obtained by a niobium deposition on a thin glass by a 
photolithographic process) placed just before the focus of the lens system. The 
two slits are separated by  100  $\mu$m  and have a width of 10 $\mu$m. They lay 
in a plane orthogonal to the incident laser beam and are orthogonal to the table 
plane (see fig.2, where the x-axis is parallel to the pump beam and the x-y plane is parallel to the optical bench). The orthogonality to the UV laser has been checked by looking at the diffraction and interference pattern of the laser by the double slit. 

Two single photon detectors are placed at 1.21 m and at 1.5 m 
from the slits after an interferential filter at 702 
nm, whose full width at half height is 4 nm, and a lens of 6 mm diameter and 25.4 mm focal length. As a preliminary step, we have also evaluated the efficiency of the detection apparatus (including losses into the crystal, filters and lenses)
by using the method described in Ref. \cite{JMO}, the  result is about $ 30 \%$.

The output signals from the detectors are routed  to a two channel counter, in 
order to have the number of events on a single channel, and to a  Time to 
Amplitude Converter (TAC) circuit, followed by a single channel analyser, for 
selecting and counting the coincidence events. 

In order to check that the two degenerate  photons crossed two different  slits, we have alternatively closed one of them by means of two sharp blades positioned on a micromovimentation, leaving the other opened; correspondingly, the coincidence peak disappeared and the coincidence signal dropped to background level. Furthermore, the signal on the related detector dropped as well, confirming the correct position of the double slit. 

 \vskip 1cm

\centerline{ \bf Calculation of the coincidence pattern predicted by Quantum Mechanics }
\vskip 0.5cm

As discussed in Ref.\cite{Mandel}, a satisfactory description of the PDC light is given by the wave function:
\be
| \Psi \rangle = | vac \rangle + \int d \omega_i d \omega_s \Phi(\omega _i,\omega _s) | \omega _i \rangle \omega _ s \rangle
\ee

In the Fraunhofer region, after selection with narrow band interferential filters (centered at the idler $\omega _i$ and signal $\omega _s$ frequencies respectively),  the diffracted field is described by:
\bea
\Phi(\omega _i,\omega _s) = g(\theta _1, \theta _i^A ) g(\theta _2, \theta _i^B) e^{-i (  {k}_A {r}_{A1} +  {k}_B {r}_{B2})} + g(\theta _2, \theta _i^A ) g(\theta _1, \theta _i^B) e^{-i (  {k}_A {r}_{A2} +  {k}_B {r}_{B1})} 
\label{phi}
\eea
where ${k}_{A}$ and ${k}_{B}$ are the wave vectors of the photon $A$ (idler) and $B$ (signal) respectively, $ {r}_{ai}$ is the vector from the slit $a$ (A or B) to the detector $i$ (1 or 2) (see Fig.2). $\theta _{1,2}$ is the diffraction angle of the photon observed by detector 1 or 2, $\theta_i^a$ the incidence angle of the photon on the slit $a$ (A or B).
\be 
g( \theta , \theta _i^a)  = { sin ( k w /2 ( sin (\theta) - sin (\theta _i^a)) 
\over 
k w /2 ( sin (\theta) - sin (\theta _i^a))}
\ee
takes into account diffraction (k is the wave vector, s the slits separation, w 
the slit's width).
The coincidence pattern that follows from Eq. \ref{phi} is:
\bea
& C(\theta_1,\theta_2) = |\Phi(\omega _i,\omega _s)|^2 = &  g(\theta _1, \theta _i^A )^2 g(\theta _2, \theta 
_i^B)^2 + g(\theta _2, \theta _i^A )^2 g(\theta _1, \theta _i^B)^2 + \cr
& & 2 g(\theta _1, \theta _i^A ) g(\theta _2, \theta _i^B) g(\theta _2, 
\theta _i^A ) g(\theta _1, \theta _i^B) cos [ k s (sin \theta _1 - sin \theta _2)]
\label{c}
\eea

The expected coincidence pattern in terms of the position of the two photo-detectors (for photons with a $2 ^o$ incidence angle) is shown in Fig.3.

In Fig.4 we show the section obtained from the previous figure when the second detector is positioned at $-1$ cm (here and in the following the positions are relative to the symmetry axis of the double slit, with a minus sign looking at the crystal from the left). The diffraction peak clearly appears modulated by the interference. 
The same graph as Fig.4 but with the second detector positioned at $-5.5 $ cm is reported in Fig.6: now a smaller interference is predicted respect to the previous one, but a larger coincidence signal is predicted when the two detector are in the same semiplane, therefore this configuration is well suited for a realisation of the experiment suggested in Ref. \cite{4,5,6} (see the following paragraph). 

We have also taken into account the effect of the non perfect monochromaticity of the PDC radiation by calculating the convolution of Eq. \ref{c} with a gaussian transfer function describing the effect of an interferential filter. We  have also introduced a small angular dispersion (1 nm/rad) of the photon pairs. The results of such a simulation show that for   a filter with a $4$ nm FWHM (corresponding to the one used in the experiment) no substantial effect appears (a detailed discussion of these effects can be found in Ref. \cite{newrad}).

\vskip 1cm

\centerline{ \bf Summary of dBB calculation of Ref.\cite{4,5,6}}
\vskip 0.5cm

Let us now briefly summarise (simplifying a bit) the results of Ref. \cite{4,5,6} concerning the dBB prediction for our double slit experiment.

Using the wave function \ref{phi} we can calculate the Bohmian velocities $\vec{v}_i = \frac{\vec{j}_i}{\Psi^* \Psi}$ of particles $i=1$ and $i=2$ (where $\vec{j}_i$ is  the current of particle $i$).
The result of this simple calculation implies that

\begin{eqnarray}
v_{1y} + v_{2y} = 0
\end{eqnarray}
This implies that

\begin{eqnarray}
y_1(t) + y_2(t) = y_1(0) + y_2(0)
\end{eqnarray}
Thus, if the initial positions of the two particles are chosen to be
symmetrical about the line of symmetry ($y = 0$), i.e., if $y_1(0) + y_2(0)
= 0$, we must have

\begin{eqnarray}
y_1(t) + y_2(t) = 0  \label{eq:5}
\end{eqnarray}
for all times, i.e., the trajectories will always be symmetrical about this
line. This implies that the two particles can, in fact, never cross the line of symmetry, in our case represented by the median axis of the double slit, and being observed in the same semiplane (For a precise explicit calculation of trajectories see Ref. \cite{5} where it is also discussed that the difference between SQM and dBB is related to fact that the last can be non-ergodic).
This is the main source of incompatibility between dBB and SQM according to Ref. \cite{4,5,6}, result that we will test in the following.

\vskip 1cm

\centerline{ \bf Experimental results}
\vskip 0.5cm

In order to scan the coincidence pattern, we have kept one detector fixed to a specific position and moved the second one. 
 
As the signal is rather low, a long acquisition time is required. This implies some problems since the power of the laser drifts over an acquisition time of several days (as we have directly verified). Also the crystal, pumped with a high power UV laser, slowly deteriorates. This effect  has been clearly observed by monitoring the observed signal on the fixed detector. 
In order to compensate this effect, we plot the average of the ratio of coincidence signal (after background subtraction) over the signal of the fixed detector multiplied for the average of the fixed detector signal. The background to coincidences is evaluated shifting the delay between the start and stop TAC inputs of 16 ns. In this way we collect the accidental coincidences far from the coincidence peak. The acquisition window for coincidences is set at $2.5$ ns.   

As a first comparison of theoretical predictions of Eq. \ref{c} with our data, in Fig. 5 we report the ones (with 10 acquisitions of one hour for each point) obtained when the first detector scans the diffraction pattern, while the second is positioned at $-1$ cm from the symmetry axis. The iris in front of the first detector is of 2 mm. Even if the data have large uncertainties there is a good indication of the fourth-order interference: the interference pattern predicted by SQM fits the data with a reduced $\chi ^ 2$  of $0.9$. By comparison, a linear fit (absence of interference) gives $\chi ^ 2 =12.6$ (with 5 degrees of freedom) and is therefore rejected with a $5 \%$ confidence level \footnote{ On the other hand we have checked that, as expected, the single channel signal does not show any variation in the same region: the measured ratio between the mobile and the fixed detector is essentially constant (within uncertainties) in this region.}.

In Fig.6 we report the swept  with a larger iris (6 mm) scanning the whole diffraction peak. The data are obtained by averaging  7 points of 30' acquisition each. The fixed detector now is placed at $-5.5$ cm from the symmetry axis.
The pattern predicted by SQM significantly agrees with the data. A clear coincidence signal is observed also when the two detectors are placed in the same semiplane respect to the double slit symmetry axis and a small signal, 41 $\pm$ 14 coincidences per hour with  17 acquisitions of one hour, is even observed in correspondence to the second diffraction peak (the area without data between the two peaks is due to the superposition of the two photodetectors). 

This last result is at variance with the  dBB prediction for coincidences calculated by \cite{4,5,6},  where the coincidence signal is predicted to be strictly zero when the two detectors are in the same semiplane with respect to the double slit symmetry axis (this configuration was purposely chosen since it has the largest coincidence signal for this case), as discussed in the previous paragraph. In particular, when the centre of the lens of the first detector is placed -1.7 cm after  the median symmetry axis  of the two slits (recall, the minus means to the left of the symmetry axis looking towards the crystal) and the second detector is kept at -5.5 cm, with 35 acquisitions of 30' each we obtained 78 $\pm$ 10 coincidences per 30 minutes after background subtraction, ruling out a null result at nearly eight standard deviations.
Thus, if  this theoretical prediction will be confirmed, this experiment poses a strong constraint on the validity of de Broglie-Bohm theory, which represents the most successful example of a non-local  hidden variable theory \footnote{Local hidden variable theories can be tested using Bell inequalities \cite{Bell}. Many experiments performed up to now \cite{bellexp} have substantially confirmed SQM, although some doubts remain due to their small detection efficiency  \cite{santos}.}.

\vskip 0.7cm
\centerline{ \bf Conclusions}
\vskip 0.5cm

In conclusion, we have realised a double slit experiment where two identical photons produced in type I PDC are sent each to a well defined slit at an identical time. 
Our data clearly show a good agreement with Quantum Mechanics predictions. By contrast, our data contradict the predictions made in  Ref. \cite{4,5,6} for de Broglie- Bohm theory, stating that no coincidences should be observed when detectors are in the same semiplane respect to the symmetry axis of the double slit. Thus, if the theoretical predictions of Ref. \cite{4,5,6} will be confirmed, our results represent a first negative test of de Broglie-Bohm theory. \footnote{This allows us to exclude also some non-conventional quantum calculation model based on special versions of the dBB theory \cite{Valentini}. Incidentally, we would like to acknowledge some previous experiments addressed to test versions of dBB theory where the empty wave has physical effects \cite{ew}.}

\vskip 0.5cm

{\bf Acknowledgments}

We acknowledge support of the Italian minister of research and of ASI under contract LONO 500172. We thank P. Ghose, A.S.Majumdar and G. Introzzi for useful discussions.  We thank R. Steni for the realisation of the double slit. 
\vskip 3cm


\newpage
\centerline{ \bf Figures Captions}
\vskip 0.5cm 
Fig.1 The experimental apparatus. A pump laser at 351 nm generates parametric 
down conversion of type I in a lithium-iodate crystal. Conjugated photons at 702 
nm are sent to a double-slit by a system of two plane-convex lenses in such a way that each photon of the pair crosses a well defined slit. A first photodetector 
is placed at  1.21 m,  a second one at 1.5 m from the slit. Each single 
photon detector (D) follows  an interferential filter at 702 nm (IF) and 
a lens (L) of 6 mm diameter and 25.4 mm focal length. Signals from the detectors are sent to a Time Amplitude Converter and then to the acquisition system (multi-channel analyser and counters). 
\vskip 0.3cm
Fig.2 Reference system. Two photons with wave vector k cross  the slits A and B of width $w=10 \mu m$ separated by $s=100 \mu m$ and are detected by the photodetectors P and Q. The x-axis is parallel to the pump beam and the x-y plane is parallel to the optical bench. 
\vskip 0.3cm  

Fig. 3 Tridimensional plot of coincidences pattern (in arbitrary units) as a function of the positions of the two photo-detectors.
\vskip 0.3cm  

Fig. 4  Plot of coincidences pattern (in arbitrary units) as a function of the positions of the first photo-detector when the second one is kept fixed at $-1$ cm from the symmetry axis.

 \vskip 0.3cm 
Fig.5 Coincidences data (with a 2 mm iris) compared with quantum mechanics predictions (solid curve).
 On the x-axis we report the position of the first detector with respect to the median symmetry axis of the double slit.
 The second detector is positioned at -0.01 m (out of scale). The leftmost region of the data is inaccessible since the two detectors overlap, while on the right, a rather flat behavior for coincidences is predicted.

\vskip 0.3cm  
Fig.6  Coincidences data (with a 6 mm iris) compared with quantum mechanics predictions (solid curve).
 On the x-axis we report the position of the first detector with respect to the median symmetry axis of the double slit.
 The second detector is kept at -0.055 m. 
The x errors bars represent the width of the lens before the detector. 
A correction for laser power  fluctuations is included.
\vskip 0.3cm  

\vskip 0.3cm  

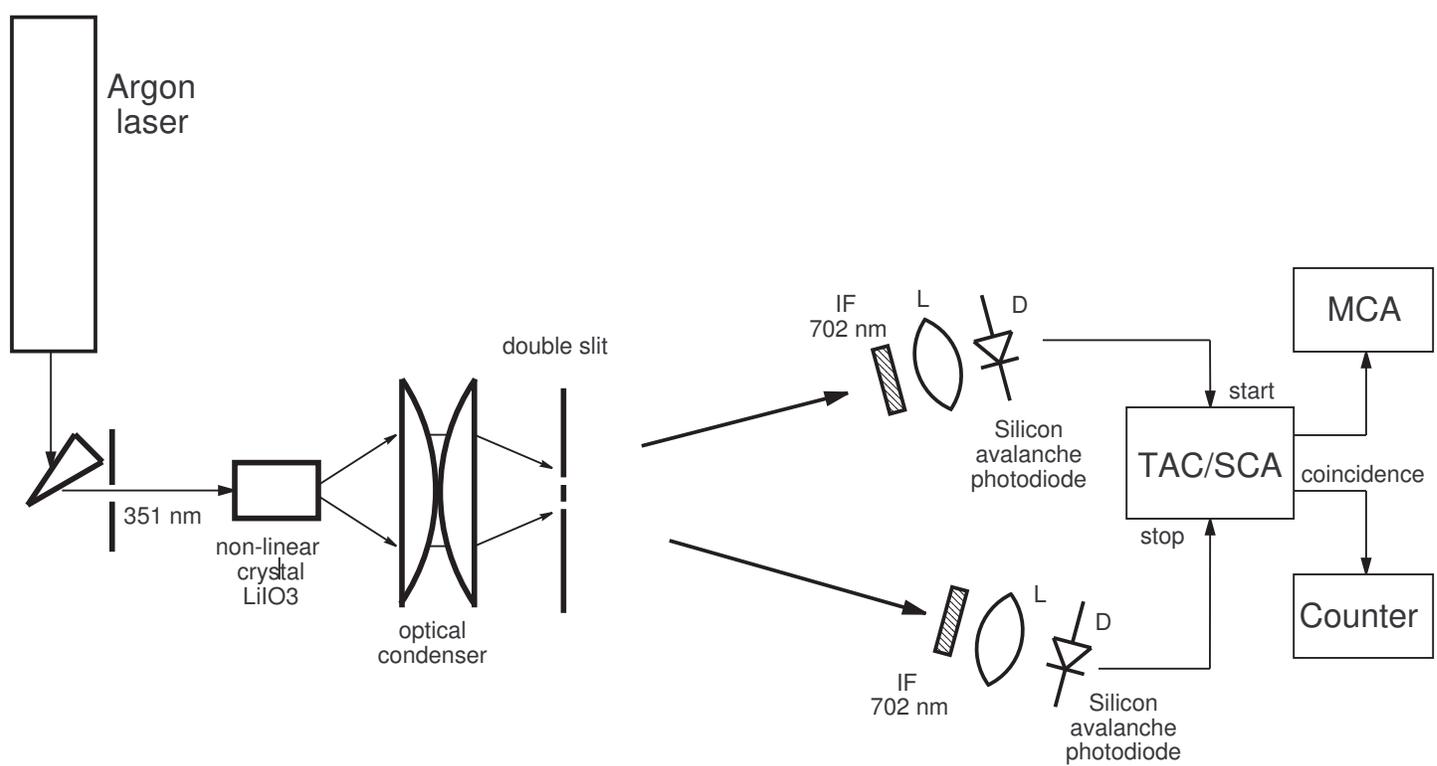
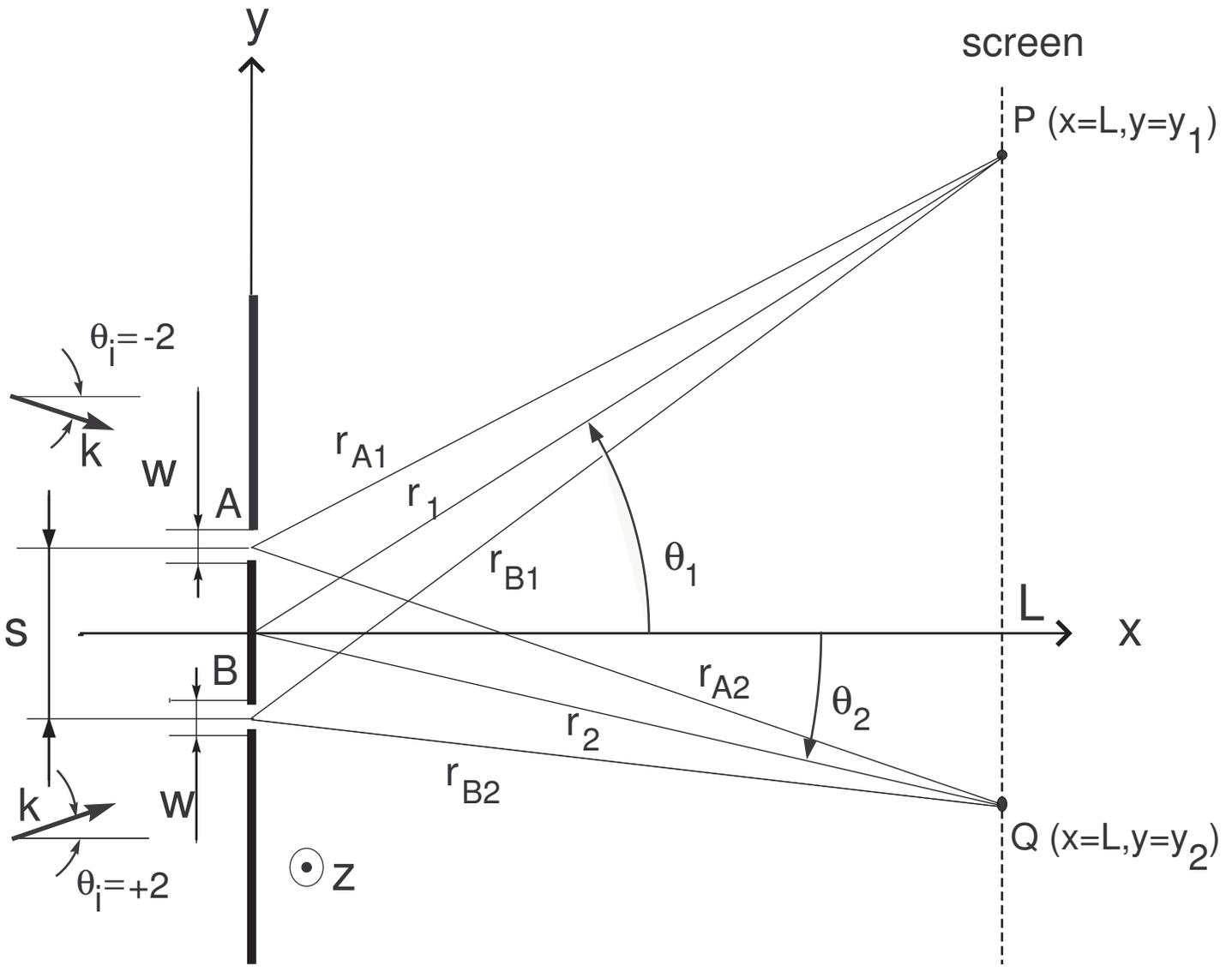
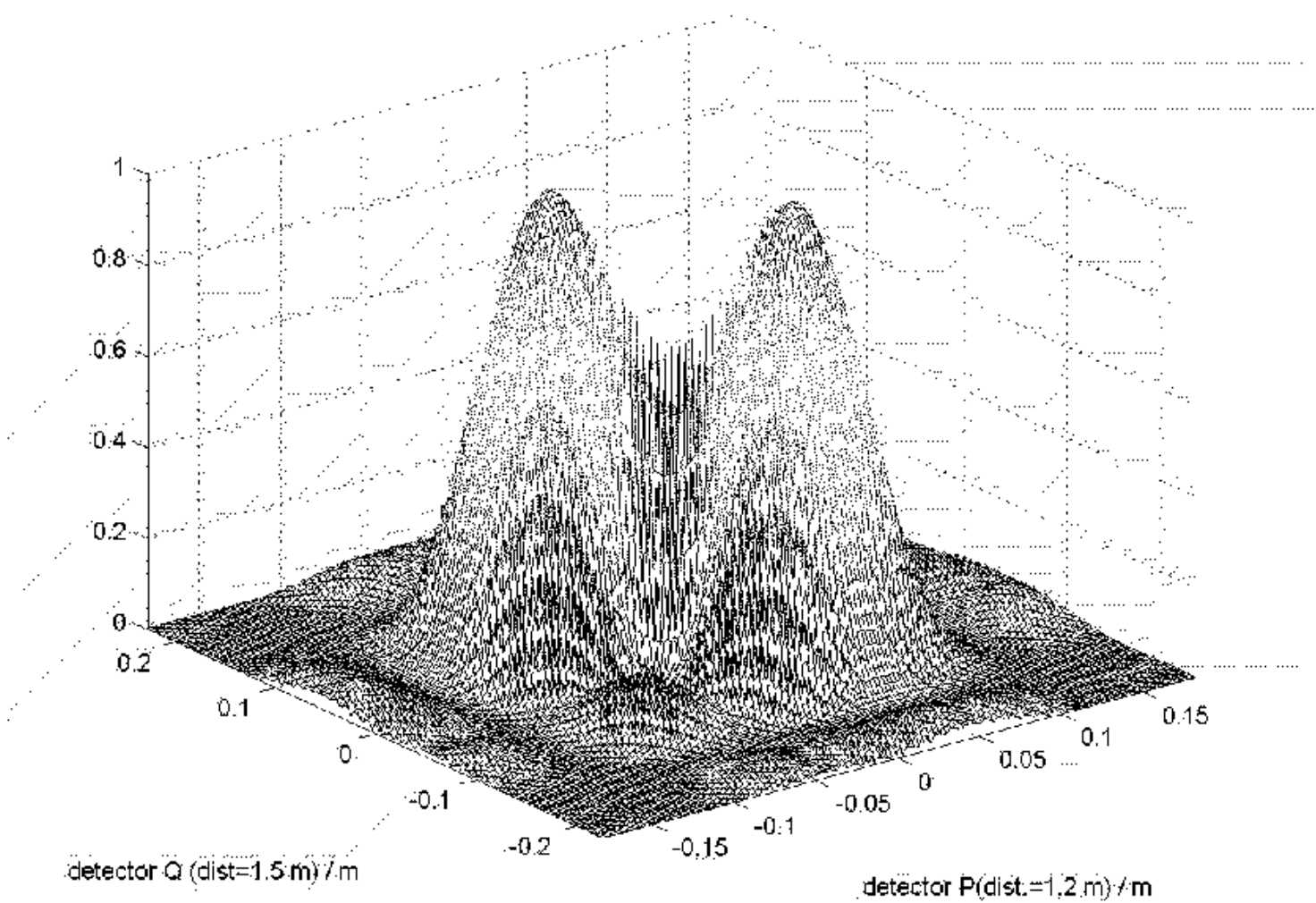
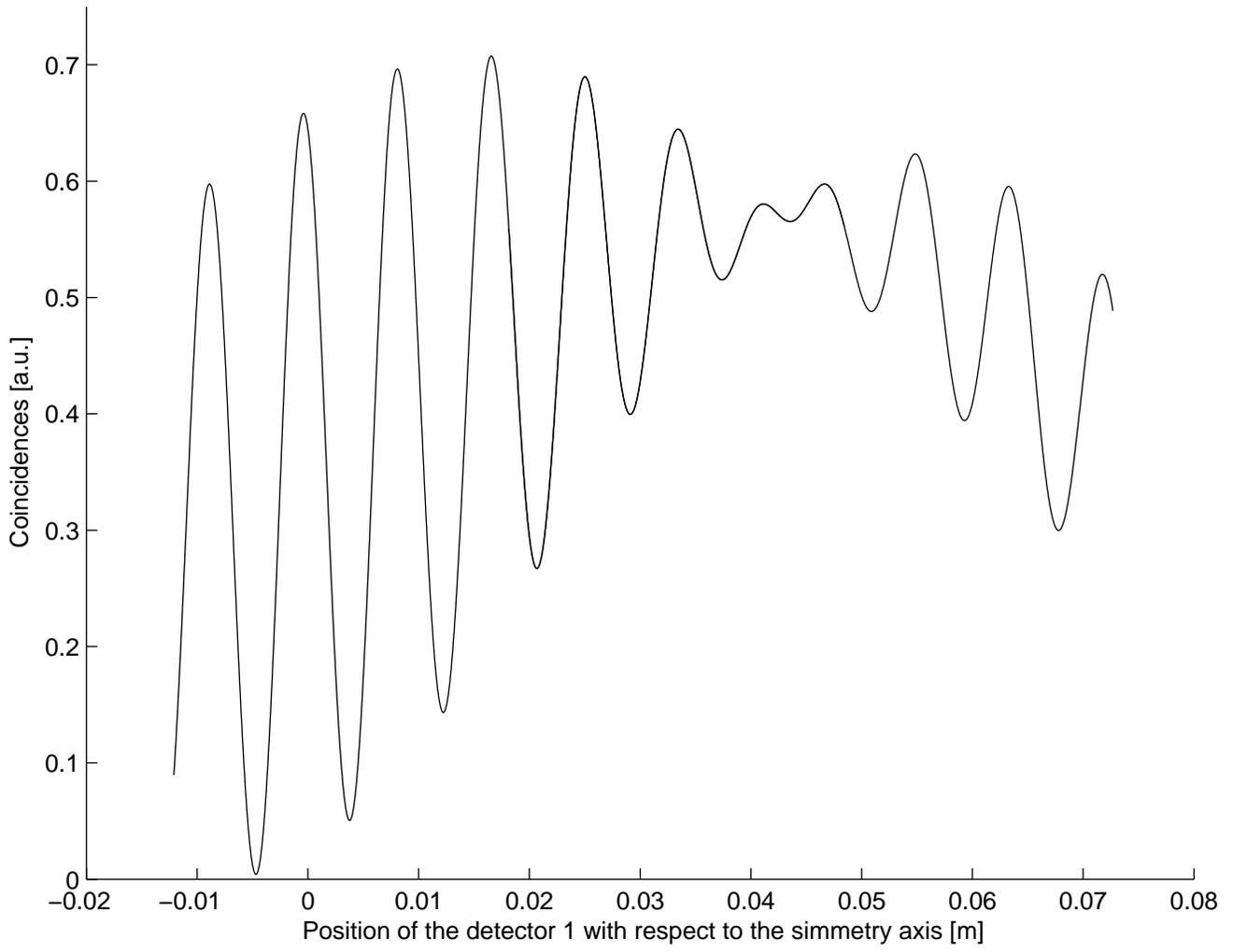
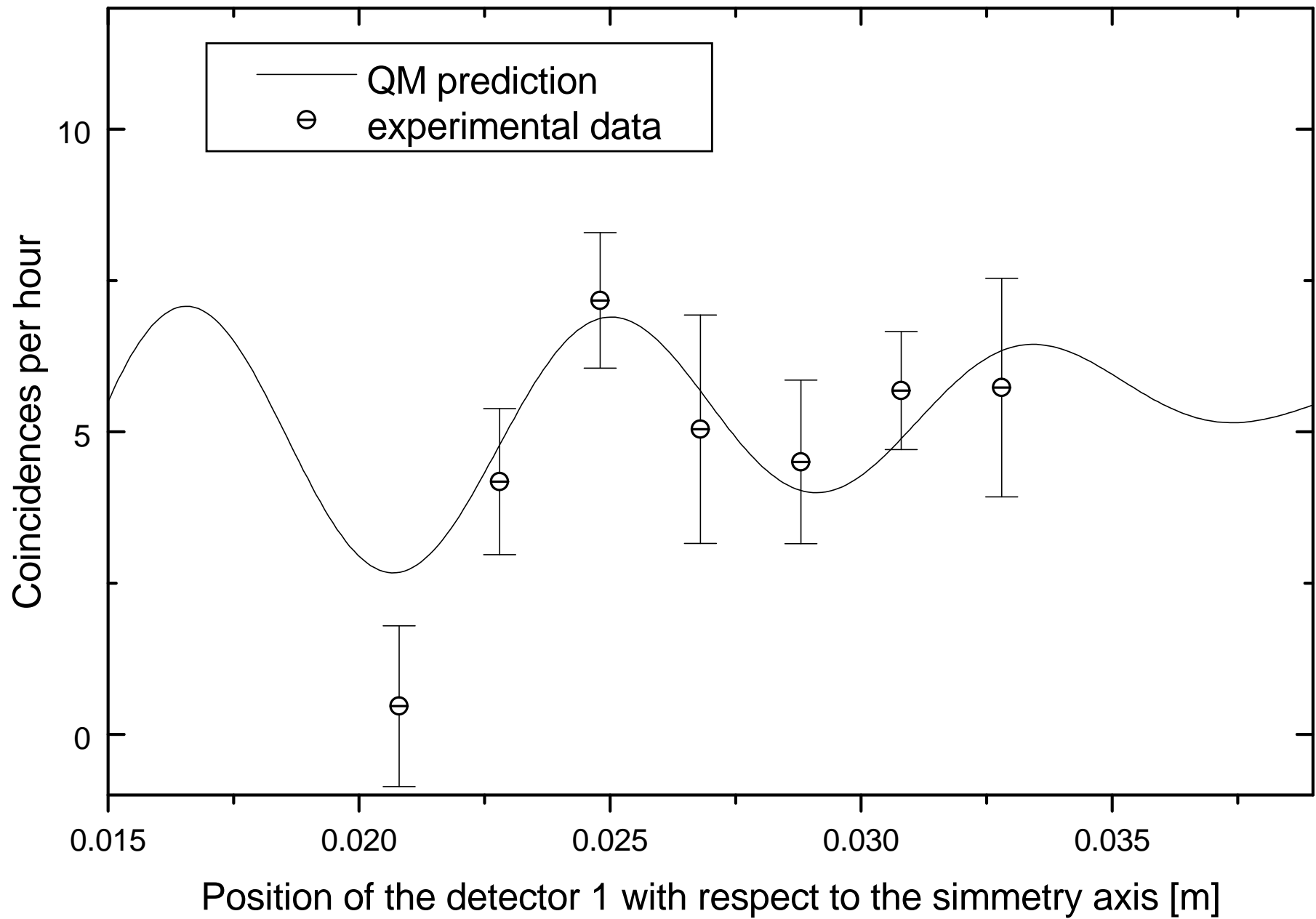
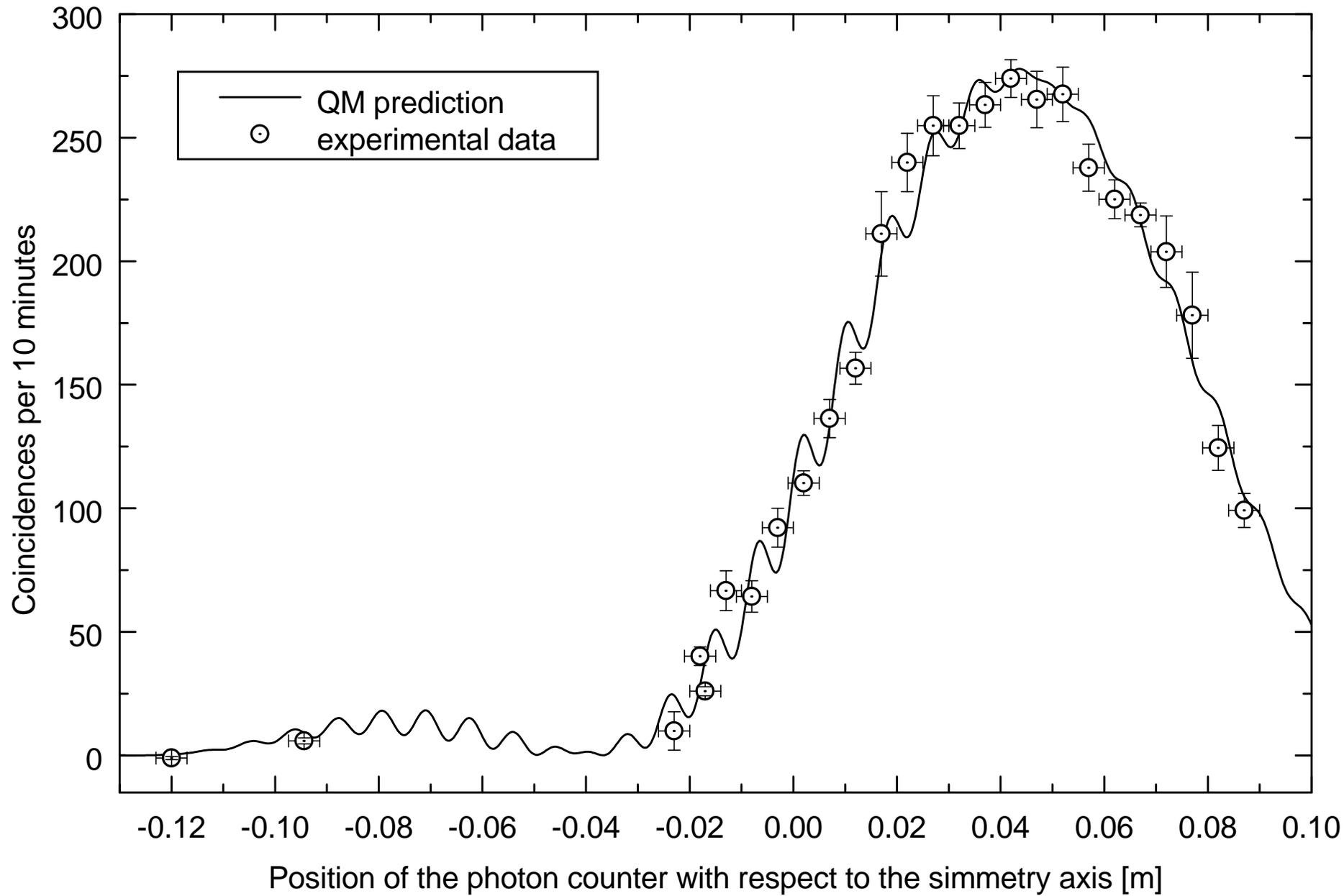
\end{document}